\documentclass[newstyle,twocolumn,preprint,proceedings]{rmaa}

\suppressfulladdresses 

\usepackage{paralist}

\usepackage{psfrag,color}

\SetYear{2006}
\SetConfTitle{Circumstellar Media and Late Stages of Massive Stellar Evolution}

\title{Polarized Line Profiles as Diagnostics of Circumstellar Geometry in Type IIn Supernovae}

\author{
  Jennifer L. Hoffman\altaffilmark{1,2}}

\altaffiltext{1}{Department of Astronomy, UC Berkeley.}
\altaffiltext{2}{NSF Astronomy \& Astrophysics Postdoctoral Fellow.}

\shortauthor{Hoffman}
\shorttitle{Polarized Lines in Type IIn SNe}

 \fulladdresses{
 \item Jennifer L. Hoffman: Department of Astronomy, University of California
at Berkeley, 601 Campbell Hall, Berkeley, CA 94720-3411, USA 
(jhoffman@astro.berkeley.edu).}

\listofauthors{J. L. Hoffman}
\indexauthor{Hoffman, J. L.}

\abstract{
Supernovae of type IIn possess spectral signatures that indicate an intense 
interaction between the supernova ejecta and surrounding dense circumstellar 
material cast off by the star in pre-explosion mass-loss episodes. Studying 
this interaction can yield clues to the nature of Type IIn progenitors and 
their mass loss history. In particular, polarization spectra of Type IIn's 
show complex line polarization and position angle features that arise from a 
combination of geometrical and optical effects. I have constructed a Monte 
Carlo code that simulates the transfer of the H$\alpha$ line through 
circumstellar shells with various geometrical configurations and optical 
characteristics. The superposition of broad and narrow line components 
produced in different regions of the circumstellar environment and modified 
by electron and line scattering, hydrogen absorption, thermal 
emission, and geometrical and viewing angle effects gives rise to a variety 
of polarized line shapes in the model spectra. Comparison of these results 
with recent high-quality spectropolarimetric observations of Type IIn 
supernovae suggests that a model ``shock" region between the supernova 
photosphere and the circumstellar shell is necessary to produce the narrow 
polarized emission features at the rest wavelength of H$\alpha$ seen in some IIn's. 
Further model results point toward other features in the polarized line profile 
that can be used to constrain the characteristics of the circumstellar material
in these intriguing objects. The code's usefulness will be extended by the 
treatment of Doppler effects due to expansion of the circumstellar scattering 
region, such as those that characterize the polarized H$\alpha$ profiles of the 
Type IIn SN 1997eg.
}

\addkeyword{Circumstellar Matter}
\addkeyword{Methods: Numerical}
\addkeyword{Polarization}
\addkeyword{Radiative Transfer}
\addkeyword{Stars: Mass loss}
\addkeyword{Supernovae}

\begin{document}
\maketitle

\section{Introduction}
\label{sec:intro}

Our understanding of supernovae (SNe) has broadened in recent years to include  
the recognition that both the thermonuclear and the core-collapse types of  
these stellar explosions are inherently aspherical phenomena 
\citep[e.g.,][]{koz05,bur06}.
Observations of net continuum polarization in both  
supernova (SN) types have provided key evidence of the intrinsic asymmetry of  
SN ejecta \citep[e.g.,][]{wang01,leon06}. Many SNe also display  
line polarization features in addition to broadband continuum polarization;  
these line effects are often more complex than simple depolarization by complete
scattering redistribution, and they can provide specific clues to the nature of 
SN ejecta and their surrounding circumstellar media 
\citep[e.g.,][]{wang04,leon05}. However, line polarization can be produced 
by a combination of many different optical and geometrical effects, so its
interpretation is not straightforward. Detailed radiative transfer modeling
\citep[as in][]{kas03} is often the best way to understand the effects that
give rise to polarized lines in SN spectra.  In the case of Type IIn
\citep[``narrow-line";][]{fil97} supernovae, the situation is complicated by
the presence of circumstellar material (CSM) surrounding the SN ejecta that
becomes
excited by the UV and X-ray photons from the SN explosion. 
Line polarization signatures in Type IIn SNe are superpositions of  
those arising from the ejecta and those arising from the CSM. Disentangling the 
two can be difficult but worthwhile, as it allows us to probe the nature and  
structure of the CSM to an extent not possible with  
spectroscopy alone \citep{leon00a,wang01}. Since the CSM of a
Type IIn supernova most likely represents material ejected by the progenitor
star, studying its geometrical and optical characteristics provides a link to the  
mass-loss episodes and stellar winds of massive stars in their late stages of  
evolution.

\section{Code}
\label{sec:code}

I have developed a Monte Carlo radiative transfer code called {\it SLIP} (for  
\underline{S}upernova \underline{Li}ne \underline{P}olarization) that simulates the ways 
polarized line profiles are created in Type IIn supernovae.~{\it SLIP} uses  
the three-dimensional spherical polar grid structure described by 
\citet{ww02}; it tracks virtual photons as they arise from a model SN  
photosphere and scatter in a circumstellar density distribution with  
wavelength-dependent emission, absorption, and scattering characteristics.  
Similar codes have enjoyed success in analyzing related scenarios such as hot  
star envelopes with aspherical wind geometries \citep{har00} and ejecta-hole  
configurations in SNe Type Ia \citep{kas04}. Unlike most previous codes  
that treat SN line polarization \citep[e.g.,][]{hoef95}, {\it SLIP} does not  
assume that line scattering is depolarizing; it also does not rely on the Sobolev
approximation, but instead performs full radiative transfer in regions of high 
optical depth. It is thus able to probe in detail the polarized line profiles 
that may arise from interaction with the circumstellar  material. Another 
advantage to this method is that it can simulate emission not only
from a central source but also from extended regions such as the warm CSM 
(see below).
However, the code is still in the early stages of development, and does not  
include any Doppler effects from the expanding circumstellar material; this  
limits the extent to which we can compare model outputs to observed line  
profiles, but the stationary case is a useful first approximation, especially  
in cases of low CSM velocity.
 
In the models presented here, a finite spherical source of photons at the  
center of the grid represents the ``photosphere" of the SN ejecta, while two 
scattering regions  surrounding the ejecta represent the warm, stationary CSM 
and a shock-heated region interior to the CSM, formed by its interaction with 
high-velocity SN material. Initially unpolarized photons are emitted from the  
surface of the photosphere with the synthetic H$\alpha$ P Cygni profile from a  
Type IIP supernova, produced with the {\it PHOENIX} stellar atmosphere  
code \citep{hau99}. In the hot shock region, I assume only a narrow  
line is emitted, with a line width of $< 80$ km/s; this is similar to the widths of  
the narrow hydrogen lines in the Type IIn SN 1997eg (7--40 km/s) observed by  
\citet{sal02}. I do not directly simulate heating of the CSM by emission from 
the supernova, but rather choose a CSM temperature and emit  
photons from the volume of the region with the expected thermal continuum and  
line spectra of hydrogen, assuming Case B LTE \citep{ost89}. All photons  
then scatter within the CSM via electron scattering; the optical depth of the  
scattering region is also chosen as an input parameter. I also implement 
wavelength-dependent free-free and free-bound absorption effects as in 
\citet{wood96}.  Photons within the Doppler core of the line \citep{lang99} 
experience an additional bound-bound opacity; those absorbed by H atoms in
this way are subsequently re-emitted coherently and isotropically. 
  
When the code is run, {\it SLIP} emits photons sequentially from the 
photosphere, shock region, and CSM and follows each  
until it becomes absorbed or exits the system; at each scattering event, the 
code updates the photon's Stokes parameters. Photons that exit the model system 
are binned by outgoing angle and the Stokes parameters are summed appropriately 
in each bin. Thus each simulation produces a full three-dimensional model 
whose H$\alpha$ flux and polarization spectra can be ``viewed" from any desired
direction.

\begin{figure*}
  \begin{center}
    \includegraphics*[width=\columnwidth]{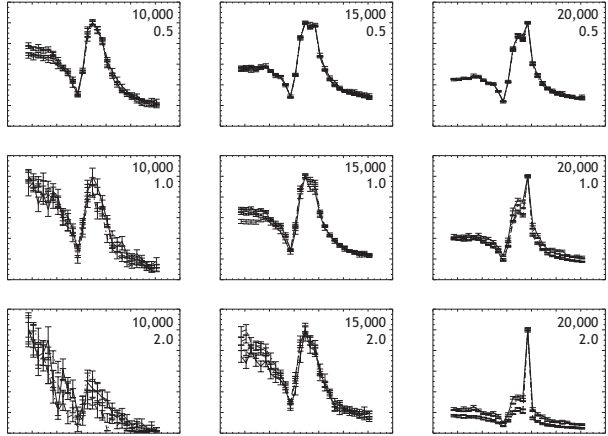}
  \end{center}
  \caption{Representative sample of the model flux grid described in \S\ref{sec:results}. 
All models shown here have ellipsoidal circumstellar density distributions; none includes 
emission from the shock region. In the individual panel labels, the upper number indicates the
CSM temperature in Kelvin and the lower indicates the CSM optical depth. Each panel spans a
wavelength range from 5800 \AA~to 7200 \AA; each spectrum is normalized to its own peak flux.
Four profiles are plotted in each panel, representing viewing angles of $3^{\circ}$, 
$35^{\circ}$, $66^{\circ}$, and $89^{\circ}$.  Model A (Table~\ref{tab:modpars}) is depicted 
in the middle right panel.\label{fig:grid}}
\end{figure*}

\section{Model Results}
\label{sec:results}

I have created a grid of 144 models spanning two CSM geometries (ellipsoids
and toroids of similar sizes); CSM optical depths from 0.5 to 2; CSM luminosities 
from  1--20\% of the photospheric luminosities; and CSM temperatures from 10,000 K to  
20,000 K. Emission from the shock region was either included at 10\% of the
photospheric luminosity or excluded completely. In order to keep computing 
times reasonable, these models included only $1.8 \times 10^{7}$ photons each,
enough to  
build up good signal in the flux spectrum but not the polarization spectrum.
Since the goal was to match general features in both the flux and polarized
flux, I used this grid to constrain parameter space for more 
computationally-intensive code runs including polarization. Figure~\ref{fig:grid} 
depicts results for nine
representative models in the flux grid; it shows that significant differences
in the line profiles arise for even small variations in optical depth and
temperature of the scattering region.
 
\begin{figure}[!t]
  \begin{center}
    \includegraphics*[width=\columnwidth]{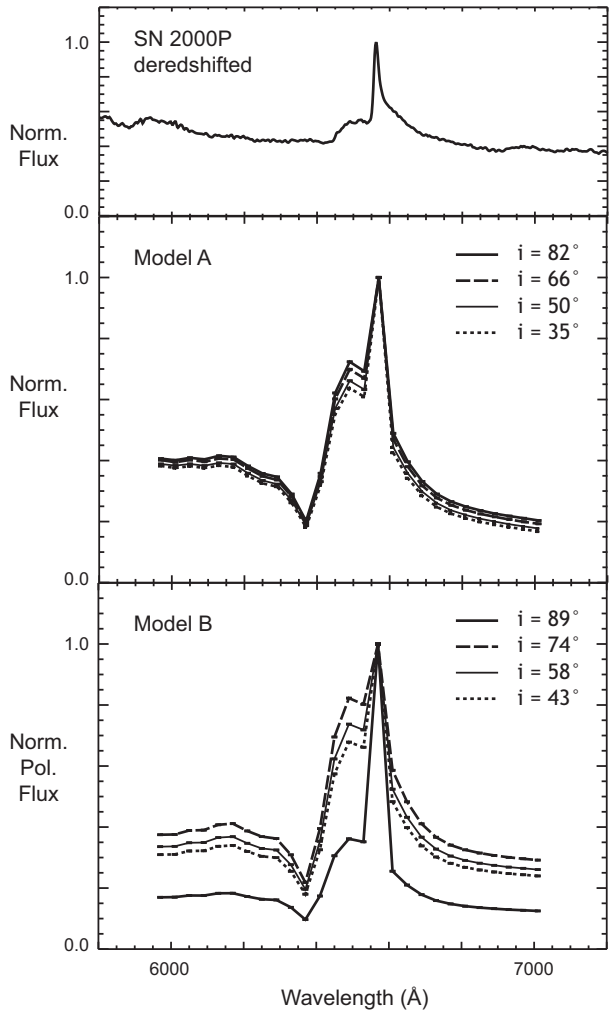}
  \end{center}
  \caption{({\it top}) De-redshifted H$\alpha$ total flux profile of SN 2000P 
at day 13 post-discovery (A. V. Filippenko 2004, priv. comm.). ({\it middle}) 
Simulated H$\alpha$ profile arising from Model A (Table~\ref{tab:modpars}) at 
varying inclination angles from the polar axis. ({\it bottom}) As in the middle 
panel, but for Model B. All line profiles have been normalized to 1 at the rest
wavelength of 6563 \AA. \label{fig:flx}}
\end{figure}

With the flux grid complete, I compared the simulated H$\alpha$ profiles with 
observed H$\alpha$ profiles of Type IIn supernovae 
to determine which simulations to repeat for better signal in the polarized
flux. This process is still underway, and the results will be published in an
upcoming contribution. Here I present two representative simulations, Model A
and Model B, that produce similar H$\alpha$ flux spectra but have quite
different polarization behavior. Each included $1.6 \times 10^{9}$ photons,
divided among 768 processors of the Seaborg parallel computing facility at the
National Energy Research Scientific Computing Center (NERSC) at the Lawrence
Berkeley Laboratory. Table~\ref{tab:modpars} compares the 
parameters of these two
models, while Figure~\ref{fig:flx} compares their H$\alpha$ line profiles at a range of
viewing angles with the profile observed for SN 2000P (A. V. Filippenko 2004, 
private communication) at day 13 post-discovery.
Both models can reproduce the general observed line shape of a narrow emission
``spike" superposed on a broad base (I note that due to the {\it SLIP} code's
limitation to stationary scattering regions, the width of the broad line in these
models arises solely from the input IIP spectrum arising from the model SN
photosphere). While the profiles produced by Model A (the ellipsoid) are nearly
completely degenerate in viewing angle, Model B (the toroid) shows a
significant variation with viewing angle, particularly in the strength of the
``spike" relative to the broad base. Examination of the other models in the 
grid suggests that the differences between these two families of H$\alpha$
profiles are mainly due to the difference in CSM geometry between these two 
models.

\begin{figure}[!t]
  \begin{center}
    \includegraphics*[width=\columnwidth]{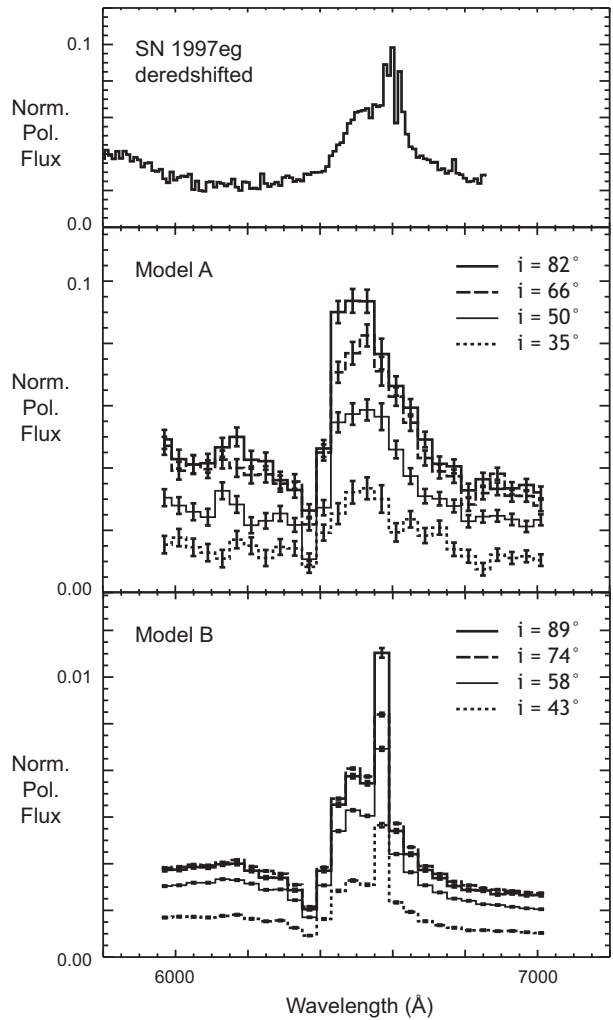}
  \end{center}
  \caption{({\it top}) De-redshifted H$\alpha$ polarized flux 
profile of SN 1997eg at day 44 post-discovery \citep{hof06}. ({\it middle}) 
Simulated H$\alpha$ polarized flux profile arising from Model A 
(Table~\ref{tab:modpars}) at varying inclination angles from the polar axis. 
({\it bottom}) As in the middle panel, but for Model B. All polarized line 
profiles have been normalized to 1 at the rest wavelength of 6563 \AA. \label{fig:pflx}}
\end{figure}

\begin{table}
  \caption{Representative Model Parameters} \label{tab:modpars}
  \begin{center}
  \begin{tabular}{lrr}\hline\hline
    Parameter & \multicolumn{1}{c}{Model A} & \multicolumn{1}{c}{Model B} \\
    \hline
    CSM geometry         & ellipsoid & toroid \\
    CSM optical depth    & 1.0       & 2.0 \\
    $L_{CSM}/L_{phot}$   & 0.01      & 0.1 \\
    $L_{shock}/L_{phot}$ & 0.0       & 0.1 \\
    CSM temperature      & 20,000 K  & 15,000K \\
    \hline\hline
  \end{tabular}
  \end{center}
\end{table}

In Figure~\ref{fig:pflx} ~I present the H$\alpha$ line profiles of Models A and B in 
polarized flux and compare them with that of the Type IIn SN 1997eg 
\citep{leon00b,hof06} at 44 days post-discovery. Recall 
that polarized flux is percent polarization multiplied by total flux; these 
profiles thus represent the spectra of the scattered light in each model. In 
polarized light the H$\alpha$ profiles look quite different than in direct 
light.
The degeneracies that characterized Model A in direct light (Figure~\ref{fig:flx}) have 
been lifted, raising the possibility of using polarized line profiles to 
diagnose inclination angle in cases of known geometry. The two models now show 
line profiles quite distinct from each other; in particular, Model A produces 
no narrow ``spike" at the rest wavelength in the polarized flux, while Model B 
preserves the spike. Examination of the other models in the flux grid suggests 
that this difference is due not to the difference in geometrical structure of 
the CSM between the two models, but rather to the presence of shock emission in 
Model B. Evidently the diffuse thermal emission from the CSM is sufficient to 
create a narrow spike in direct light, but in polarized light this feature
requires a more directional source of narrow-line photons (the shock region). 
Photons from this region are more likely to scatter and become polarized while 
traversing the CSM than are photons that arise within its volume. However,
Model B does not match the observed magnitude of polarization in SN 1997eg, most
likely due to its larger CSM optical depth.

I continue to investigate these model results to pinpoint further diagnostics 
for geometry, temperature, and optical depth of the circumstellar material and 
the presence of a narrow-line ``shock" region in Type IIn ejecta. More detailed
analysis will be published in an upcoming contribution.

\section{Observed H$\alpha$ Line Profiles of SN 1997eg}
\label{sec:97eg}

\begin{figure}[!t]
  \begin{center}
    \includegraphics{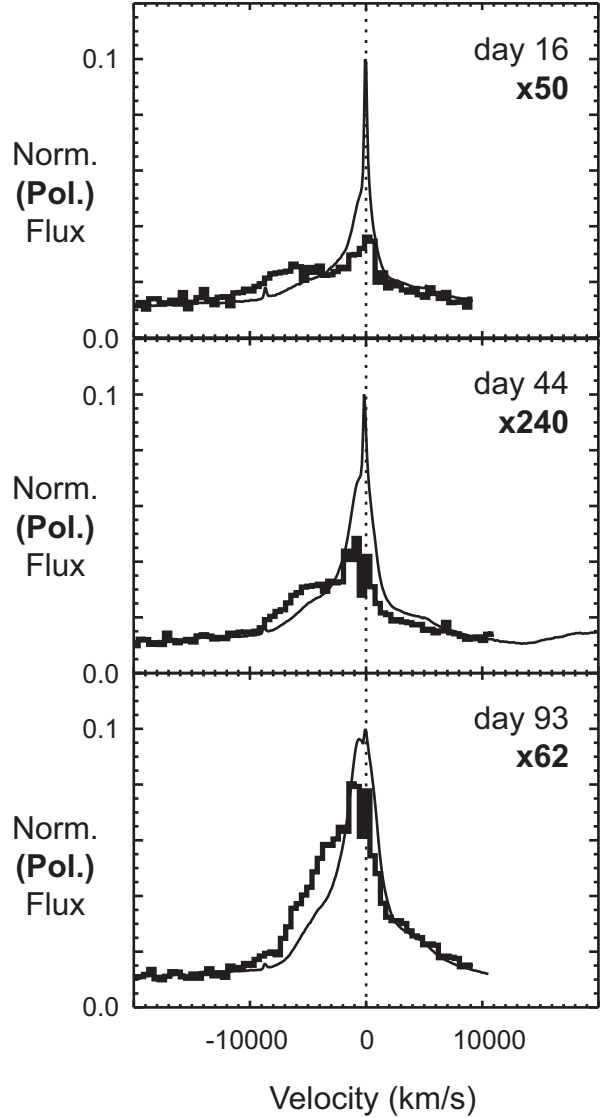}
  \end{center}
  \caption{Line profiles of the H$\alpha$ line of SN 1997eg in total flux 
({\it narrow smooth lines}) and polarized flux ({\it thick binned lines}) at 
each of the three epochs of spectropolarimetry (days 16, 44, and 93 
post-discovery). Total flux spectra have been normalized to their respective 
line peaks. Each polarized flux spectrum has been binned to a resolution of
10 \AA~and multiplied by the same normalizing factor as its corresponding 
total flux spectrum, then by an additional factor shown in each frame to 
facilitate direct comparison of the line shapes. \label{fig:widths}}
\end{figure}

\begin{figure}[!t]
  \begin{center}
    \includegraphics{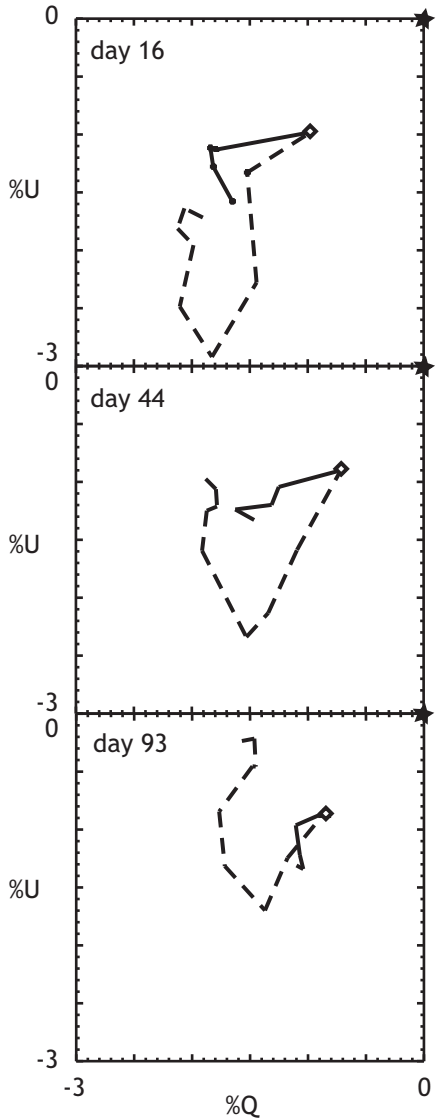}
  \end{center}
  \caption{Polarization profiles of the H$\alpha$ emission line of SN 1997eg 
in the {\it q--u} plane for each of the three epochs of spectropolarimetry 
(days 16, 44, and 93 post-discovery). Data have been binned to a resolution of 
50 \AA~for clarity. Dashed lines represent negative (blueshifted) velocities; 
solid lines represent positive (redshifted) velocities. In each frame, the 
rest wavelength is shown with an open diamond and the origin of the plot is 
in the upper right corner. \label{fig:qu}}
\end{figure}

As mentioned in \S\ref{sec:code}, my current {\it SLIP} code has the 
limitation of treating only stationary scattering regions, when in fact one 
expects velocity effects to be quite prominent contributors to line profiles 
in polarized light. Here I present two examples of line polarization effects 
in a Type IIn supernova that cannot yet be reproduced by the code but provide 
key information regarding the geometry of the circumstellar material. Along with 
collaborators at San Diego State University and UC Berkeley, I have studied 
the polarization spectrum of SN 1997eg (whose H$\alpha$ polarized flux profile 
is shown in Figure~\ref{fig:pflx}); our full analysis will appear in \citet{hof07}.

Figure~\ref{fig:widths} compares the H$\alpha$ line profiles of SN 1997eg in total flux 
(nearly all direct light) and scattered light at day 16, day 44, and day 93 
post-discovery. Although the continuum polarized flux is only about 2\% of the 
total 
at each epoch, I have normalized both spectra to the same scale for comparison 
of the line profiles. At all epochs the polarized lines are broader than the 
unpolarized lines. This implies first that the scattering region has a different 
geometry from that of the broad-line emission region, and second that the scattering
region is expanding at a higher velocity than the emission region. The fact that 
the polarized lines are 
broader only in the blue wing suggests the redshifted side of the scattering 
region may be self-occulted (or perhaps occulted by the SN ejecta) from our line of 
sight. If we postulate a flattened toroidal or disk-like geometry for the scattering 
region, this result can help place limits on the spatial inclination of the CSM
configuration.

In Figure~\ref{fig:qu} I plot the polarized H$\alpha$ lines of the three epochs in 
{\it q--u} space, a technique that allows visualization of all polarimetric 
information at once. Not only does the magnitude of polarization change across 
the H$\alpha$ line at all epochs in SN 1997eg, but the position angle changes 
as well, in a manner that creates closed ``loops" in the {\it q--u} plane 
(distinct from ``knots," which characterize a constant polarization signal with 
wavelength, and from straight lines, which arise from simple envelope expansion 
or line depolarization). This implies that the scattering region polarizing the
H$\alpha$ line has a different orientation than the one polarizing the continuum 
light (presumably the SN ejecta). In particular, the ``loop" shape implies that
the symmetry axes of the two regions are different, and that the CSM occults
the SN ejecta in such a way as to create an asymmetry in the Stokes parameters 
across the line center. In the models of \citet{kas03}, such {\it q--u} 
loops are general features of two-axis systems; I postulate a toroidal geometry
such as that shown in these authors' Figures 14 and 15.

I note that the preceding results are both independent of interstellar polarization 
effects, for which the observed data have not been corrected. Similar complex 
behavior in the H$\beta$ and \ion{He}{i} $\lambda$5876 lines in the SN 1997eg 
spectrum is discussed fully in the upcoming article. The results combine to 
suggest that SN 1997eg is characterized by ellipsoidal ejecta that 
polarize the continuum light and a flattened, disk-like CSM exterior to the 
ejecta that polarizes the hydrogen lines. The key conclusion is that the 
ejecta and the CSM have different axes of symmetry.

Multi-axis systems are becoming recognized as quite common in 
mass-loss scenarios. P Cyg \citep{nor01,mea01}, $\eta$ Car \citep{smith06}, 
and VY CMa \citep{hum05} are examples of 
massive stellar systems in which the circumstellar material shows evidence
for multiple mass-loss episodes along different axes. 
Our observations of polarized line profiles in SN 1997eg suggest that the 
geometry of an asymmetric SN explosion may, in turn, be unrelated to the geometry 
of the progenitor's stellar wind or its mass eruptions. Continued refinements to 
the {\it SLIP} radiative transfer code will allow me to
construct more detailed models of the CSM surrounding SN 1997eg and related
objects, as well as further probing the nature both of Type II SN explosions and the
eruptions that characterize their progenitor stars.

This research was supported by an NSF Astronomy \& Astrophysics Postdoctoral
Fellowship, AST-0302123, and by the National Energy Research Scientific Computing
Center, US DOE Contract \#DE-AC03-76SF00098. I thank my primary collaborators
Alex Filippenko at UC Berkeley, Peter Nugent at LBL, and Doug Leonard at
San Diego State University, for their invaluable contributions.

\end{document}